\documentclass[%
 reprint,
 amsmath,amssymb,
 aps,
]{revtex4-2}

\usepackage{graphicx}
\usepackage{dcolumn}
\usepackage{bm}
\usepackage{mathtools}
\usepackage{color}

\begin{document}

\preprint{APS/123-QED}

\title{Ornstein-Uhlenbeck information particle: A new candidate of active agent}

\author{Xin Song}
\thanks{These authors contribute equally to this work.}
\affiliation{Department of Physics, Wenzhou University, Wenzhou, Zhejiang 325035, China}
\affiliation{Zhejiang Key Laboratory of Soft Matter Biomedical Materials, Wenzhou Institute, University of Chinese Academy of Sciences, Wenzhou, Zhejiang 325001, China}

\author{Xiji Shao}
\thanks{These authors contribute equally to this work.}
\affiliation{Department of physics, School of Intelligent Engineering, Shaoguan University, Shaoguan, Guangdong 512005, China}

\author{Yanwen Zhu}
\thanks{These authors contribute equally to this work.}
\affiliation{Department of Physics, Wenzhou University, Wenzhou, Zhejiang 325035, China}
\affiliation{Zhejiang Key Laboratory of Soft Matter Biomedical Materials, Wenzhou Institute, University of Chinese Academy of Sciences, Wenzhou, Zhejiang 325001, China}

\author{Cheng Yang}
\affiliation{School of Physics and Electronic Information, Mianyang Teachers’ College, Mianyang 621000, China}

\author{Linli He}\email{Corresponding author: linlihe@wzu.edu.cn}
\affiliation{Department of Physics, Wenzhou University, Wenzhou, Zhejiang 325035, China}

\author{Shigeyuki Komura}\email{Corresponding author: komura@wiucas.ac.cn}
\affiliation{Zhejiang Key Laboratory of Soft Matter Biomedical Materials, Wenzhou Institute, University of Chinese Academy of Sciences, Wenzhou, Zhejiang 325001, China} 

\author{Zhanglin Hou}\email{Corresponding author: zl\_hou@tju.edu.cn}
\affiliation{Zhejiang Key Laboratory of Soft Matter Biomedical Materials, Wenzhou Institute, University of Chinese Academy of Sciences, Wenzhou, Zhejiang 325001, China}


\begin{abstract}
An information particle can acquire active-like motion through transforming the information entropy into effective self-propulsion velocity/force using the attached information engine. We consider an underdamped Brownian particle additionally driven by either a constant self-propulsion force or an information engine using Ornstein-Uhlenbeck (OU) bath feedback control, such particles are called self-propelled particle (SPP) or OU information particle (OUIP). Compared to the widely-investigated SPP, the OUIP shows a significant different dynamical pattern, including two types of moving mode: a slow-speed diffusion mode and a high-speed traveling mode. The specific evolution of OUIP can be adjusted flexibly between such two modes through the inertial effect, thus acquiring a rich and non-trivial motion behavior. By tuning the strength of fluctuation of the OU bath, a wide range of net velocity can be achieved for OUIP. We highlight that OUIP could be an exceptional candidate for active agent.
\end{abstract}

\maketitle

\section{\label{sec:introduction}Introduction}
Active motion of self-propelled units breaks the time-reversal symmetry, bringing the system out of equilibrium, which makes the system more complicated, but provides more opportunities for innovation in fundamental research and applications~\cite{bechinger2016active}. Physically, microscale active units acquire their self-propelling activities through various sources. For example, bacteria swim or crawl powered by filaments~\cite{hu2024multiflagellate, merz2000pilus, wadhwa2022bacterial}, synthetic swimmers move through the generated gradient field~\cite{wang2013small, moran2017phoretic, kitahata2013spontaneous, loffler2023new}, motions of passive particles are driven by external (force, magnetic, electric, optical, thermal, acoustic, or potential) fields~\cite{romanczuk2012active, astumian2002brownian, kato2022active, mandal2018magnetic, gangwal2008induced, ghosh2009controlled, jiang2010active, wang2012autonomous, platten2006soret}, and continuous mechanical energy input~\cite{narayan2007long, komatsu2015roles}, etc.

An information swimmer has recently been proposed that self-propulsion can be achieved by transforming information entropy into mechanical work through an information engine~\cite{ huang2020information, hou2025ornstein}. As supposed by Szilard~\cite{szilard1929entropieverminderung, parrondo2015thermodynamics}, system can extract work from attached thermal reservoir using information through measurement and feedback controls, which has been widely realized in theoretical and experimental efforts~\cite{koski2014experimental, bengtsson2018quantum, ribezzi2019large, aydiner2021space, paneru2020colloidal, malgaretti2022szilard, toyabe2010experimental, saha2023information, paneru2018lossless} and recently introduced in Brownian system~\cite{huang2020information, hou2025ornstein, paneru2018lossless, park2016optimal, paneru2018optimal}. The realization of information swimmer depends on three operations: measurement, information storage, and feedback control. More specifically, in a Brownian system, an underdamped particle acquires a finite net velocity by periodically measuring its velocity and adjusting either its friction coefficient~\cite{huang2020information} or the persistence time of attached Ornstein-Uhlenbeck (OU) noise~\cite{hou2025ornstein}.

Active particles can provide great promising applications both for understanding non-equilibrium phenomena and guidance of realistic challenges based on their moving 
features.~\cite{bechinger2016active} For example, self-propelled particles performing persistent random walks are used to mimic the run-and-tumble dynamics of bacteria~\cite{romanczuk2012active, solon2015active}, a light-dependent activity-tunable Janus particle emerges as an exceptional candidate for clogging or unclogging of microchannels~\cite{buttinoni2012active,dressaire2017clogging,baldovin2023control}. On the other hand, the most attractive aspect of active matter systems, which has gained widespread attention, arises from their spontaneously emergent phenomena, such as motility-induced phase separation (MIPS)~\cite{cates2015motility}, spontaneous velocity alignment~\cite{caprini2020spontaneous,vicsek1995novel}, and the ability to imitate various complicated natural entities, e.g., flocking and swarming of birds/insects, and predator-prey behavior~\cite{mandal2018magnetic}.

Unfortunately, realistic units always exhibit a non-trivial and complicated behavior. Taking living organisms for example, bacteria execute their motion according to environmental conditions (nutrient/chemical gradient~\cite{berg1972chemotaxis,matthaus2009coli}, psl trails~\cite{zhao2013psl}, etc.), density-guided self-organization behavior of mussels~\cite{van2008experimental,liu2013phase}, and so on. As a consequence, a simple active agent's behavior is not enough to simulate their behaviors, the one who has a higher controllability, operability, and richer movement features is a better candidate.

Based on previous works on information swimmer~\cite{huang2020information, hou2025ornstein}, we extend the one-dimensional model to a two-dimensional (2D) system through polar coordinates, making it as an active agent. Specifically, we choose the OU noise feedback model~\cite{hou2025ornstein} to mimic the internal control of the active agent itself (OU information particle (OUIP), see the model below). To better clarify the characteristics of the proposed OUIP, we further consider a self-propelled particle (SPP) driven by a constant self-propulsion force for comparison, which was previously discussed by Suvendu Mandal et al.~\cite{mandal2019motility}. In general, we propose a 2D active underdamped colloidal particle whose motion is either driven by an additional constant force or tuned by an information engine through an attached OU noise. In the past few decades, SPP has received extensive attention due to its success in simulating/searching the collective behavior or emergent exotic phenomenon/effect of realistic units. Here, we introduce the aforementioned model for comparing SPP and OUIP. Through numerical investigation, we show that though both SPP and OUIP can acquire an average net velocity along their orientation direction, their motions show quite different features. A further comparison demonstrates that OUIP can show a richer tunable motion pattern, should be regarded as an exceptional candidate for active agent.

\section{\label{sec:model}Model and simulations}
We consider a 2D underdamped Brownian particle additionally driven by either a self-propulsion force ($\gamma_{\rm t} U_0$) or an information engine with OU noise ($u(t)$) feedback control. To better depict the activity of particle, the velocity of a particle $\boldsymbol{v}_{\rm p}$ is decomposed into a parallel component $v_{\parallel} \mathbf{e}_{\parallel}$ and a perpendicular component $v_{\perp} \mathbf{e}_{\perp}$, $\boldsymbol{v}_{\rm p} = v_{\parallel} \mathbf{e}_{\parallel} + v_{\perp} \mathbf{e}_{\perp}$, where $v_{\parallel} \mathbf{e}_{\parallel}$ is the component parallel to the orientation of the particle with $\mathbf{e}_{\parallel} = (\cos\theta, \sin\theta)$, $v_{\perp} \mathbf{e}_{\perp}$ is perpendicular to it, $\mathbf{e}_{\perp} = (-\sin\theta, \cos\theta)$, and $\theta$ the orientation of the particle. The dynamics of particle are modeled by the following equations:
\begin{gather}
m \dot{v}_{\parallel} (t) = -\gamma_{\rm t} v_{\parallel} (t) + \gamma_{\rm t} U + \sqrt{2 k_{\text{B}} T \gamma_{\rm t}} \eta_{\parallel} (t),
\label{EQ:parallelmotion}
\\
m \dot{v}_{\perp} (t) = -\gamma_{\rm t} v_{\perp} (t) + \sqrt{2 k_{\text{B}} T \gamma_{\rm t}} \eta_{\perp} (t),
\label{EQ:perpendicularmotion}
\\
I \ddot{\theta} (t) = -\gamma_{\rm r} \dot{\theta} (t) + \sqrt{2 k_{\text{B}} T \gamma_{\rm r}} \eta_{\theta} (t).
\label{EQ:rotationmotion}
\end{gather}
Here $m$, $I$, $\gamma_{\rm t}$ and $\gamma_{\rm r}$ are the mass, moment of inertia, translational and rotational friction coefficients of particle, respectively, $T$ the temperature of system, and $k_{\text{B}}$ the Boltzmann constant. $\eta_{\parallel}$, $\eta_{\perp}$ and $\eta_{\theta}$ represent Gaussian white noises with zero mean and unit variance:
\begin{equation}
    \langle \eta_i \rangle = 0, \quad \langle \eta_i (t) \eta_j  (t^{\prime}) \rangle =\delta_{ij} \delta(t-t^{\prime}),
    \label{EQ:noises}
\end{equation}
where the subscripts $i$ and $j$ represent $\parallel$, $\perp$ or $\theta$. 

In the present work, the parameter $U$ in Eq.~\ref{EQ:parallelmotion} can be either a self-propulsion speed $U_0$ or an OU noise $u(t)$. In the case of $U=U_0$, the particle is driven by a self-propulsion force as well as the white noise, and thus is regarded as SPP. Whereas $U=u(t)$, the dynamics of the particle will be controlled by the information engine through periodic measurement and feedback operations, combining the perturbation of white noise, thus we regard the particle as OUIP. For OUIP, using an information engine, we adjust the persistence time of the OU noise through dimensionless parameters $\beta_1$ and $\beta_2$ according to the measurement results (the parallel component of the velocity $v_{\parallel}$ is larger or smaller than a threshold velocity $v_0$) at every time interval $\tau_{\rm{m}}$. The specific process follows the equations
\begin{equation}
\dot{u} (t) = 
\begin{dcases}
-\frac{\beta_1 u(t)}{\tau_{\rm{a}}} + \frac{\beta_1 \sqrt{2A}}{\tau_{\rm{a}}} \zeta(t), & \quad v_{\parallel} (n \tau_{\rm{m}}) \leq v_0,
\\
-\frac{\beta_2 u(t)}{\tau_{\rm{a}}} + \frac{\beta_2 \sqrt{2A}}{\tau_{\rm{a}}} \zeta(t), & \quad v_{\parallel} (n \tau_{\rm{m}}) > v_0,
\end{dcases}
\label{EQ:OUnoisefeedback}
\end{equation}
for the time within $n \tau_{\rm{m}} < t \le (n+1) \tau_{\rm{m}}$. Here, $\tau_{\rm{a}}$ is the persistence time, $A$ is the perturbation strength, $\beta_1$ and $\beta_2$ are dimensionless parameters introduced to adjust the persistence time, and $\zeta(t)$ is zero mean and unit variance Gaussian white noise:
\begin{equation}
\langle \zeta (t) \rangle = 0, \quad \langle \zeta (t) \zeta (t^{\prime}) \rangle = \delta (t-t^{\prime}).
\label{EQ:noise_zeta}
\end{equation}

It is convenient to discuss the models using dimensionless variables and parameters. We use the relaxation time (inertial delay time) $\tau = m/ \gamma_{\rm t}$, thermal velocity $v_T = \sqrt{k_{\rm B}T/m}$, and characteristic length $l_0 = \sqrt{mk_{\rm B}T} / \gamma_{\rm t}$ to rescale all variables and parameters. The corresponding dimensionless equations are shown in Appendix~\ref{sec:dimensionless}.

In the following analysis, we will use dimensionless time $\tilde t = t / \tau$, measurement time $\tilde \tau_{\rm{m}} = \tau_{\rm{m}} / \tau$, persistence time $\tilde \tau_{\rm{a}} = \tau_{\rm{a}} / \tau$, velocities $\tilde v_{\parallel}(\tilde t) = v_{\parallel}(t)/v_T$, $\tilde U_0=U_0/v_T$, $\tilde v_0 = v_0 / v_T$, as well as $\tilde v_x (\tilde t) = v_x (t) / v_T$ in Cartesian coordinate, and the perturbation strength of OU noise $\tilde A = A \gamma_{\rm t} / (k_{\rm B}T)$. Furthermore, we introduce dimensionless mass $\tilde M =mk_{\rm B}T/(\gamma_{\rm t} \gamma_{\rm r})$ to measure the impact of inertia of the particle.

We perform single-particle dynamics simulation in a square box with periodic boundary condition. In simulations, we first run $10^3 \tau$ to ensure that the particle reaches a steady moving state, and subsequently another $5.5 \times 10^5 \tau$ is run to record the state of the particle to perform statistical analysis.

\section{\label{sec:model}Results}

\begin{figure*}[htbp]
\centering
\includegraphics[width=0.9\textwidth]{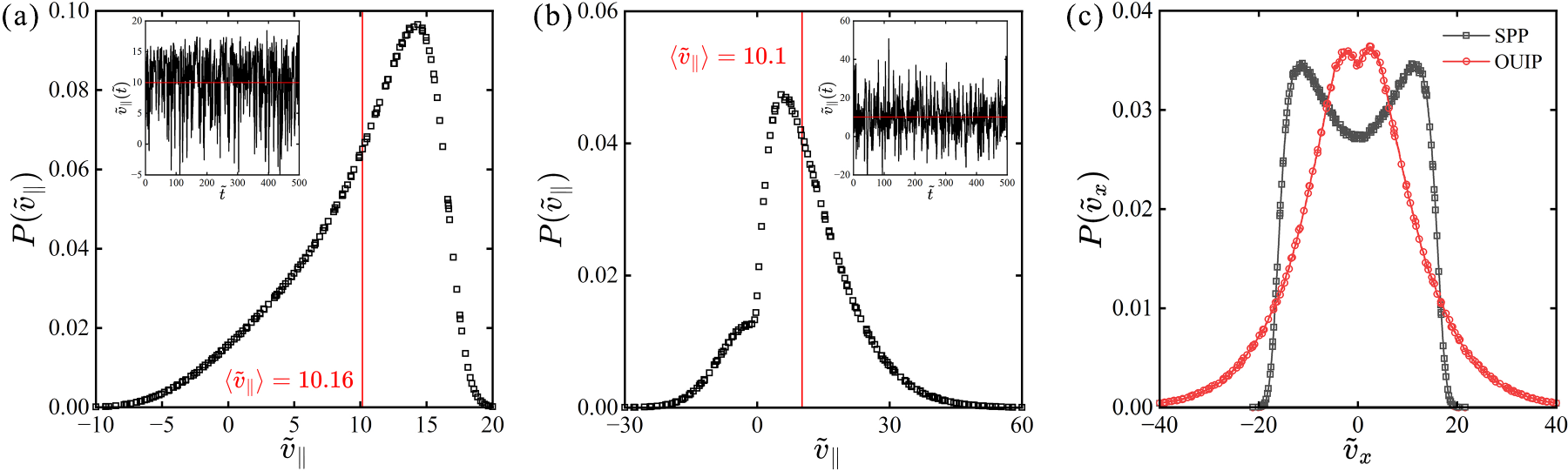}
\caption{
(a) The velocity distribution probability of SPP in the orientation direction, where $\tilde U_0 = 18$ and $\tilde M=1$. Inset: The time evolution of velocity. The red lines denote the average velocity of SPP along its orientation direction.
(b) The velocity distribution probability of OUIP along its orientation direction in polar coordinates, where $(\beta_1, \beta_2) = (10, 0.1)$, threshold velocity $\tilde v_0 = 0$, measurement time interval $\tilde \tau_{\rm{m}} = 10 \Delta \tilde t$ with $\Delta \tilde t = 0.001$, and the fixed parameters are $\tilde A = 63$, $\tilde \tau_{\rm{a}} = 1$ and $\tilde M=1$. Inset: The time evolution of velocity. The red solid lines denote the average velocity of OUIP along its orientation direction. 
(c) The distribution probability of velocity $P(\tilde v_x)$ of SPP and OUIP on $x$-axis in Cartesian coordinate system. The parameters of particles are same as in (a) and (b).
}
\label{FIG:OUIP&SPP}
\end{figure*}

Both the SPP and OUIP can acquire a finite net average velocity along their orientation directions in polar coordinates under suitable conditions. Figure \ref{FIG:OUIP&SPP} shows the velocity distributions of SPP and OUIP, where they both have a net average velocity $\langle \tilde v_{\parallel} \rangle \sim 10$ in their orientation directions. Although their curves have a similar average value, they exhibit different distributions, indicating that the motion behavior is quite different between these two particles. The distribution curve of SPP has a peak at $\tilde v_{\parallel} \sim 15$ (Fig.~{\ref{FIG:OUIP&SPP}}(a)), corresponding to its constant driving velocity $\tilde U_0 = 18$. The long tail on the smaller value side originates from the effect of inertia (the curve recovers a Gaussian form under the overdamped approach, e.g., the case of $\tilde M \sim 0.01$). The time evolution of velocity (inset in Fig.~\ref{FIG:OUIP&SPP}(a)) reveals that the instantaneous velocity of SPP almost holds at a high speed level with some low speed moments exist. In general, the SPP maintains a high-speed moving velocity under the constant driving force.

In contrast, though the velocity distribution curve of the OUIP is not symmetric as well, due to the information engine breaking the time-reversal symmetry of system, its profile is significantly different from the SPP case, as shown in Fig.~\ref{FIG:OUIP&SPP}(b). In OUIP, the distribution curve becomes discontinuous near the threshold velocity point $\tilde v_{\parallel}=\tilde v_0$. The pattern of its movement can be treated as a combination of two kinds of motion behaviors, including a low-speed diffusion part that sits at $\tilde v_{\parallel} \sim 0$ and a high-speed traveling part whose center is slightly higher than $\tilde v_0$. Compared with the SPP case, the instantaneous velocity of OUIP that is shown in the inset of Fig.~\ref{FIG:OUIP&SPP}(b) tends to be more unstable. It keeps moving forward at a high speed overall, accompanied by extremely fast forward and a certain probability of slow or even backward movement behaviors, while the SPP maintains a high speed forward with some kinds of disturbances. It is convenient to distinguish their different movement features from their dynamical behaviors directly, see Movies S1 and S2 for the representative movements of SPP and OUIP.

The distribution probability of velocity is also calculated in the Cartesian coordinate, and the distribution curves of SPP and OUIP on the $x$-axis are shown in Fig.~{\ref{FIG:OUIP&SPP}}(c). Consistent with the observations above, SPP and OUIP show different distribution features. The velocity of SPP exhibits a symmetric saddle form with peaks locating at the positions whose absolute values are slightly smaller than its driving velocity $\tilde U_0$, demonstrating that the high-speed travel motion is the dominant regime of SPP. However, the velocity of OUIP concentrates around $\tilde v_x = 0$ with a broad tail (a stretched exponential tail that will be discussed in detail below) that holds a finite probability value in the extreme high value window. Although SPP and OUIP have a similar average net velocity in their orientation directions, their average velocities on the $x$-axis are both $0$ in the long time limit, but show different movement behaviors. We will focus on the behavior of OUIP below and clarify its specific characteristics.

\begin{figure*}[htbp]
\centering
\includegraphics[width=0.9\textwidth]{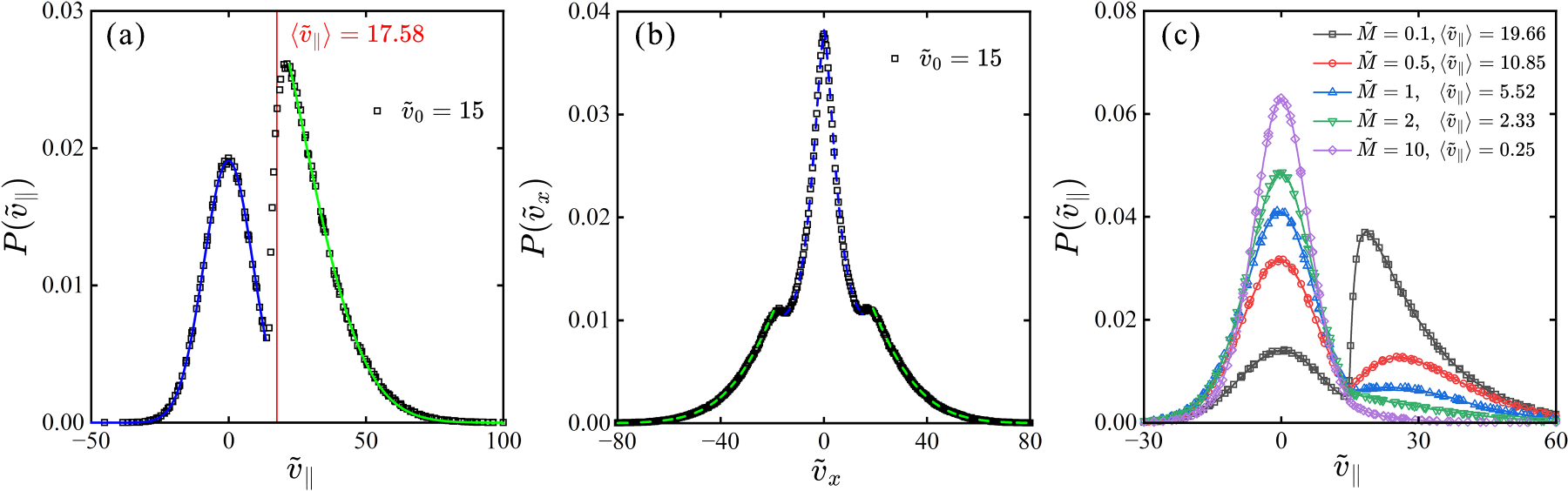}
\caption{
The motion characteristics (a-b) and inertial-effect (c) of selected representative OUIP.
(a) The velocity distribution $P(\tilde v_{\parallel})$ of OUIP in the orientation direction of particle. The red solid vertical line denotes the average velocity of OUIP along its orientation direction. The blue and green solid lines are Gaussian fittings.
(b) The velocity distribution $P(\tilde v_x)$ of OUIP on the $x$-axis of Cartesian coordinate in space. The dashed blue and green lines are stretched exponential fittings ($P(\tilde v_x) \sim \exp(-(\vert \tilde v_x/v_c \vert)^{\alpha})$ with $\alpha \sim 1.445 $ and $1.65$, respectively).
(c) The velocity distribution probability $P(\tilde v_{\parallel})$ of OUIP under various inertia effects that controlled by reduced mass $\tilde M$.
The parameters of selected OUIP are $(\beta_1, \beta_2) = (10, 0.1)$, threshold velocity $\tilde v_0 = 15$, measurement time interval $\tau_{\rm{m}} = 10 \Delta \tilde t$ with $\Delta \tilde t = 0.001$, and the fixed parameters are $\tilde A = 100$ and $\tilde \tau_{\rm{a}} = 1$. The reduce mass in (a) and (b) is fixed as $\tilde M = 0.2$.
}
\label{FIG:OUIP}
\end{figure*}

The OUIP with a selective threshold velocity of $\tilde v_0 = 15$ and reduced mass of $\tilde M = 0.2$ is further investigated. In this case, particle acquires a net average velocity of $\langle \tilde v_{\parallel} \rangle = 17.58$. Its velocity distributions are shown in Figs.~{\ref{FIG:OUIP}} (a) and (b). For $\tilde v_{\parallel}$ where the velocity is along the orientation direction, $P(\tilde v_{\parallel})$ shows two peaks. The first peak sits at $\tilde v_{\parallel} \sim 0$ and the other one locates just around $\tilde v_{\parallel} \sim 15$ (i.e., $\tilde v_0$). In general, the distribution $P(\tilde v_{\parallel})$ consists of two separated Gaussian curves, indicating that the OUIP includes two types of motion behaviors, slow motion and fast motion modes. As a consequence, the distribution of velocity on $x$-axis in Cartesian coordinate $P(\tilde v_x)$ exhibits two distinct parts (Fig.~\ref{FIG:OUIP}(b)), corresponding to the low-speed diffusing regime and the high-speed traveling regime. The distribution curve between these two regimes exhibits a significant discontinuous change around $\vert \tilde v_x \vert \sim \tilde v_0$. These separated parts in $P(\tilde v_x)$ can both be described through a stretched exponential fitting with exponent values much lower than 2, indicating that the motion of OUIP has been influenced by the OU bath and thus shows an OU noise-modulated behavior, details see the Appendix~\ref{sec:AOUP} and the discussion below.

Notably, the motion behavior of OUIP consisting of two distinguishable modes is tunable between such modes through the inertial effect of the particle, i.e., the value of $\tilde M$. This results from the competition between particle's inertia and the dissipation of ambient environment. A higher inertia will enhance the low-speed moving mode, whereas the increase of dissipation will weaken the effect of inertia and promote the high-speed mode, as shown in Fig.~\ref{FIG:OUIP}(c).

The features of OUIP can be interpreted by combining the behavior of AOUP discussed in Appendix~\ref{sec:AOUP}. The attached OU noise bath periodically switches between the $\beta_1$ and $\beta_2$ states according to the measurement results. The resulting curve should be a combination of such two states, because these states can be regarded as two independent processes. As a consequence, all the $P(\tilde v_{\parallel})$, $P(\tilde v_x)$, and motion behavior (Movie~S2) show two distinct well-separated  modes. As discussed in previous investigation~\cite{hou2025ornstein}, due to the weak and strong effect of inertia in states $\tilde v \leq \tilde v_0$ and $\tilde v > \tilde v_0$, respectively, the switching behaviors provide the chance that the weak inertia state partially converts to the strong inertia state, acquiring a finite net average velocity in the direction that the information engine works. Finally, OUIP exhibits an inertia-tunable behavior.

\begin{figure}[htbp]
\centering
\includegraphics[width=0.3\textwidth]{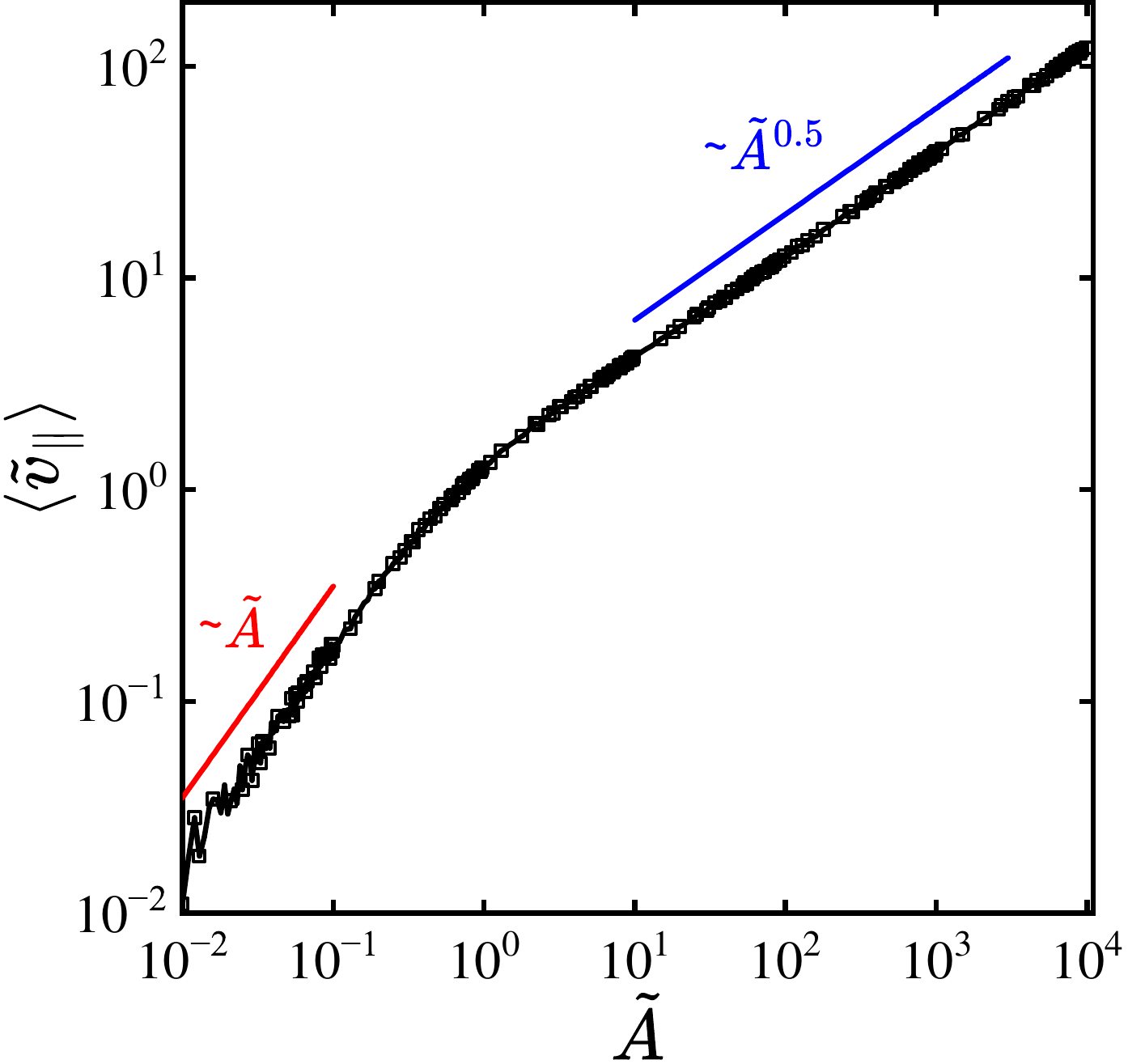}
\caption{
Obtaining net average velocity $\langle \tilde v_{\parallel} \rangle$ of OUIP by tuning the value of $\tilde A$.
Here, we choose $(\beta_1, \beta_2) = (10, 0.1)$, threshold velocity $\tilde v_0 = 0$, measurement time interval $\tilde \tau_{\rm{m}} = 10 \Delta \tilde t$ with $\Delta \tilde t = 0.001$, the fixed parameters are $\tilde \tau_{\rm{a}} = 1$ and $\tilde M = 1$.
}
\label{FIG:A-dependent-velocity}
\end{figure}

We further investigate the locomotion of OUIP. Taking the $\tilde v_0 = 0$ case for example, by adjusting the perturbation strength value $\tilde A$, OUIP can acquire a wide range of net average (effective driving) velocity $\langle \tilde v_{\parallel} \rangle$. As shown in Fig.~{\ref{FIG:A-dependent-velocity}}, the obtaining average velocity of OUIP increases monotonically with the value of $\tilde A$. In principle, the growth curve is effective temperature $T^{\ast}$ dependent, which roughly obeys $\langle \tilde v_{\parallel} \rangle \sim \sqrt{k_{\rm B}T^{\ast} / m}$ with effect temperature $T^{\ast} = T [1+ \tilde A / (1+ \tilde \tau_{\rm{a}} / \beta)]$~\cite{hou2025ornstein}. When $\tilde A \ll 1$, $\langle \tilde v_{\parallel} \rangle \approx \sqrt{k_{\rm B}T / m} \{ 1+ \tilde A / [2(1+ \tilde \tau_{\rm a} / \beta] \}  \sim \tilde A$, while for the case of $\tilde A \gg 1$, $\langle \tilde v_{\parallel} \rangle \approx \{ k_{\rm B} T \tilde A / [m (1+\tilde \tau_{\rm a} / \beta)] \}^{1/2} \sim \tilde A ^ {1/2}$. This growth behavior is independent of the inertial effect, see Fig.~\ref{FIG:S2}. As a consequence, by changing $\tilde A$ from $10^{-2}$ to $10^4$, the $\langle \tilde v_{\parallel} \rangle$ exhibits two-stage power-law-form growth behaviors. We emphasize that, in general, OUIP can achieve a wide range of activity, i.e., the effective driving velocity $\langle \tilde v_{\parallel} \rangle$, by tuning the value of $\tilde A$.

\section{\label{sec:model}Summary and discussion}
By introducing an information engine to tune the dynamical behavior of underdamped Brownian particle through the feedback of attached OU noise bath, we show that an OUIP can acquire a stable average net velocity, providing OUIP the potential application as a model particle for simulating active or living objects. Comparing with the underdamped SPP, OUIP shows a significantly different motion behavior due to their distinct self-propulsion mechanism. The activity of OUIP includes two types of moving modes, a low-speed mode and a high-speed mode. It is possible to combine and adjust these two moving modes flexibly to achieve various complex motion patterns, providing an promising application prospect of OUIP, e.g., the run-and-tumble behavior of bacteria. We stress that OUIP should be a good candidate of active agent for exploring the fundamental issues about biological physics and bioinspired engineering where self-propulsion relevant.

The properties of OUIP we discussed above provide an opportunity for OUIP to serve as a model particle of active agent. The wide range of effective self-propulsion speed $\langle \tilde v_{\parallel} \rangle$ ensures OUIP the ability to search the full space phase diagram. Moreover, the tunable non-uniform motion pattern makes the possibility for OUIP to simulate/clarify the objects with complex motion behavior or high-requirement realistic use. In underdamped SPP system, the activity of particle (the self-propulsion speed, i.e., $\tilde U$) closely relates to the characteristic time of collision event between particles, which will concern the occurrence of motility-induced phase separation (MIPS) in system~\cite{mandal2019motility}. Unlike SPP, the existence of low-speed and high-speed motion patterns in OUIP provides an unstable or multiple characteristic time scales in system, which is a promising direction worth to investigate in the future. In our model, just an additional thermal bath and an information engine are utilized to control the behavior of particle. It is also interesting to search for more complicated behavior of OUIP through multiple information engines and thermal baths' control, e.g., the achievement of chiral particle when an information engine works on the orientational motion of particle. Furthermore, it is also worth considering to design the model and investigate the feedback response based on the information measured from the environment of unit rather than particle itself~\cite{vicsek1995novel,vansaders2023informational}, which should bring much broader application in the future.

\begin{acknowledgments}
We thank Bin Zheng and Zhongqiang Xiong for useful discussion. 
This project is supported by the National Natural Science Foundation of China (Grant 22273067 to L.H.; Grant 12274098 to S.K.; and Grant 12104453 to Z.H.).
S.K. and Z.H. acknowledge the support by the Zhejiang Key Laboratory of Soft Matter Biomedical Materials (2025ZY01036 and 2025E10072).
S.K. acknowledges the support by the Scientific Research Starting Foundation of Wenzhou Institute, UCAS (Grant No. WIUCASQD2021041) and the Japan Society for the Promotion of Science (JSPS) Core-to-Core Program “Advanced core-to-core network for the physics of selforganizing 
active matter” (No. JPJSCCA20230002).
X.S. acknowledges the support by the Startup Funding from Shaoguan University (440/9900064706) and the Solid State Physics Teaching Steering Committee of the Ministry of Education Teaching Reform Project (JZW-24-GT-03).
\end{acknowledgments}

\appendix

\section{\label{sec:dimensionless}Dimensionless models}
We use the Brownian relaxation time $\tau = m / \gamma_{\rm t}$, thermal velocity $v_T = \sqrt{k_{\rm B}T/m}$, and characteristic length $l_0 = \sqrt{mk_{\rm B} T} / \gamma_{\rm t}$ to rescale the variables and parameters as $\tilde t =t / \tau$, $\tilde \tau_{\text{m}} = \tau_{\text{m}} / \tau$, $\tilde \tau_{\text{a}} = \tau_{\text{a}} / \tau$, $\tilde \tau_{\rm r} = I / (\tau \gamma_{\rm r})$, $\tilde v_{\parallel}(\tilde t) = v_{\parallel}(t)/v_T$, $\tilde v_{\perp}(\tilde t) = v_{\perp}(t)/v_T$, $\tilde U = U / v_T$, $\tilde U_0=U_0/v_T$, $\tilde u (\tilde t) = u(t) / v_T$, $\tilde v_0 = v_0 / v_T$, $\tilde{\eta}_{\parallel} (\tilde t) = \sqrt{2m/\gamma_{\rm t}} \eta_{\parallel} (t)$, $\tilde{\eta}_{\perp} (\tilde t) = \sqrt{2m/\gamma_{\rm t}} \eta_{\perp} (t)$, $\tilde \eta_\theta (\tilde t) = \tau \sqrt{2k_{\rm B} T/\gamma_{\rm r}} \eta_\theta (t)$, $\tilde{\zeta} (\tilde t) = \sqrt{2m/\gamma_{\rm t}} \zeta (t)$, and $\tilde A = A \gamma_{\rm t} / (k_{\rm B} T)$. As a consequence, the dimensionless form of the dynamics of the particle models can be rewritten as
\begin{gather}
\frac{d \tilde{v}_{\parallel} (\tilde t)}{d \tilde t} = - \tilde{v}_{\parallel} (\tilde t) + \tilde U + \tilde{\eta}_{\parallel} (\tilde t),
\label{EQSI:parallelmotion}
\\
\frac{d \tilde{v}_{\perp} (\tilde t)}{d \tilde t} = - \tilde{v}_{\perp} (\tilde t) + \tilde{\eta}_{\perp} (\tilde t),
\label{EQSI:perpendicularmotion}
\\
\tilde \tau_{\rm r} \frac{d \dot{\theta} (\tilde t)}{d \tilde t} = - \dot{\theta} (\tilde t) + \tilde{\eta}_{\theta} (\tilde t),
\label{EQSI:rotationmotion}
\end{gather}
with Gaussian white noises
\begin{gather}
\langle \tilde{\eta}_{\parallel} (\tilde t) \rangle = 0, \quad \langle \tilde{\eta}_{\parallel} (\tilde t) \tilde{\eta}_{\parallel} (\tilde t ^{\prime}) \rangle = 2 \delta (\tilde t - {\tilde t}^{\prime}),
\label{EQSI:noiseparallel}
\\
\langle \tilde{\eta}_{\perp} (\tilde t) \rangle = 0, \quad \langle \tilde{\eta}_{\perp} (\tilde t) \tilde{\eta}_{\perp} (\tilde t ^{\prime}) \rangle = 2 \delta (\tilde t - {\tilde t}^{\prime}),
\label{EQSI:noiseperpencular}
\\
\langle \tilde{\eta}_{\theta} (\tilde t) \rangle = 0, \quad \langle \tilde{\eta}_{\theta} (\tilde t) \tilde{\eta}_{\theta} (\tilde t ^{\prime}) \rangle = \frac{2mk_{\rm B} T}{\gamma_{\rm t} \gamma_{\rm r}}\delta (\tilde t - {\tilde t}^{\prime}),
\label{EQSI:noiserotation}
\end{gather}
and their cross-relationship
\begin{equation}
    \langle \tilde \eta_{\parallel} (\tilde t) \tilde \eta_{\perp} (\tilde t^{\prime}) \rangle = 
    \langle \tilde \eta_{\parallel} (\tilde t) \tilde \eta_{\theta} (\tilde t^{\prime}) \rangle = 
    \langle \tilde \eta_{\perp} (\tilde t) \tilde \eta_{\theta} (\tilde t^{\prime}) \rangle = 0.
\end{equation}
Here, we introduce reduced mass $\tilde M = mk_{\rm B}T / (\gamma_{\rm t} \gamma_{\rm r})$ to measure the impact of particle inertia.

The information engine controlled Ornstein-Uhlenbeck (OU) bath is rewritten as

\begin{equation}
\frac{d \tilde u (\tilde t)}{d \tilde t} =
\begin{dcases}
- \frac{\beta_1 \tilde u (\tilde t)}{{\tilde \tau}_{\text{a}}} + \frac{\beta_1 \sqrt{\tilde A}}{{\tilde \tau}_{\text{a}}} \tilde \zeta (\tilde t), & \quad \tilde v_{\parallel} (n {\tilde \tau}_{\text{m}}) \leq \tilde v_0, 
\\
- \frac{\beta_2 \tilde u (\tilde t)}{{\tilde \tau}_{\text{a}}} + \frac{\beta_2 \sqrt{\tilde A}}{{\tilde \tau}_{\text{a}}} \tilde \zeta (\tilde t), & \quad \tilde v_{\parallel} (n {\tilde \tau}_{\text{m}}) > \tilde v_0.
\end{dcases}
\label{EQSI:OUbathfeedback}
\end{equation}
for $n {\tilde \tau}_{\text{m}} < \tilde t \le (n+1) {\tilde \tau}_{\text{m}}$, with Gaussian white noise
\begin{equation}
\langle \tilde{\zeta} (\tilde t) \rangle = 0, \quad \langle \tilde{\zeta} (\tilde t) \tilde{\zeta} (\tilde t ^{\prime}) \rangle = 2 \delta (\tilde t - {\tilde t}^{\prime}).
\label{EQSI:noiseOUbath}
\end{equation}

\section{\label{sec:AOUP}The dynamics of active Ornstein-Uhlenbeck particle (AOUP)}
It is necessary to clarify the dynamics of underdamped Brownian particle with OU colored noise attached (active OU particle, AOUP) in two-dimensions, which is important to understand the delicate features of motion behavior of OUIP. The dynamics of AOUP are described by the Eqs.~(\ref{EQ:parallelmotion}-\ref{EQ:noises}) with $U=u(t)$. Here, $u(t)$ is the attached OU noise, which evolves through an Ornstein-Uhlenbeck process
\begin{equation}
\dot{u} (t) = -\frac{\beta u(t)}{\tau_{\rm{a}}}+\frac{\beta \sqrt{2A}}{\tau_{\rm{a}}} \zeta (t),
\label{EQSI:OUnoise}
\end{equation}
with white noise $\zeta (t)$ defined in Eq.~\ref{EQ:noise_zeta}. Its corresponding dimensionless form is
\begin{equation}
\frac{d \tilde u (\tilde t)}{d \tilde t} = -\frac{\beta \tilde u (\tilde t)}{\tilde{\tau}_{\text{a}}} + \frac{\beta \sqrt{\tilde A}}{\tilde{\tau}_{\text{a}}} \tilde \zeta (\tilde t),
\label{EQSI:OUbath}
\end{equation}
with related white noise Eq.\ref{EQSI:noiseOUbath}.

\begin{figure*}[htbp]
\centering
\includegraphics[width=0.6\textwidth]{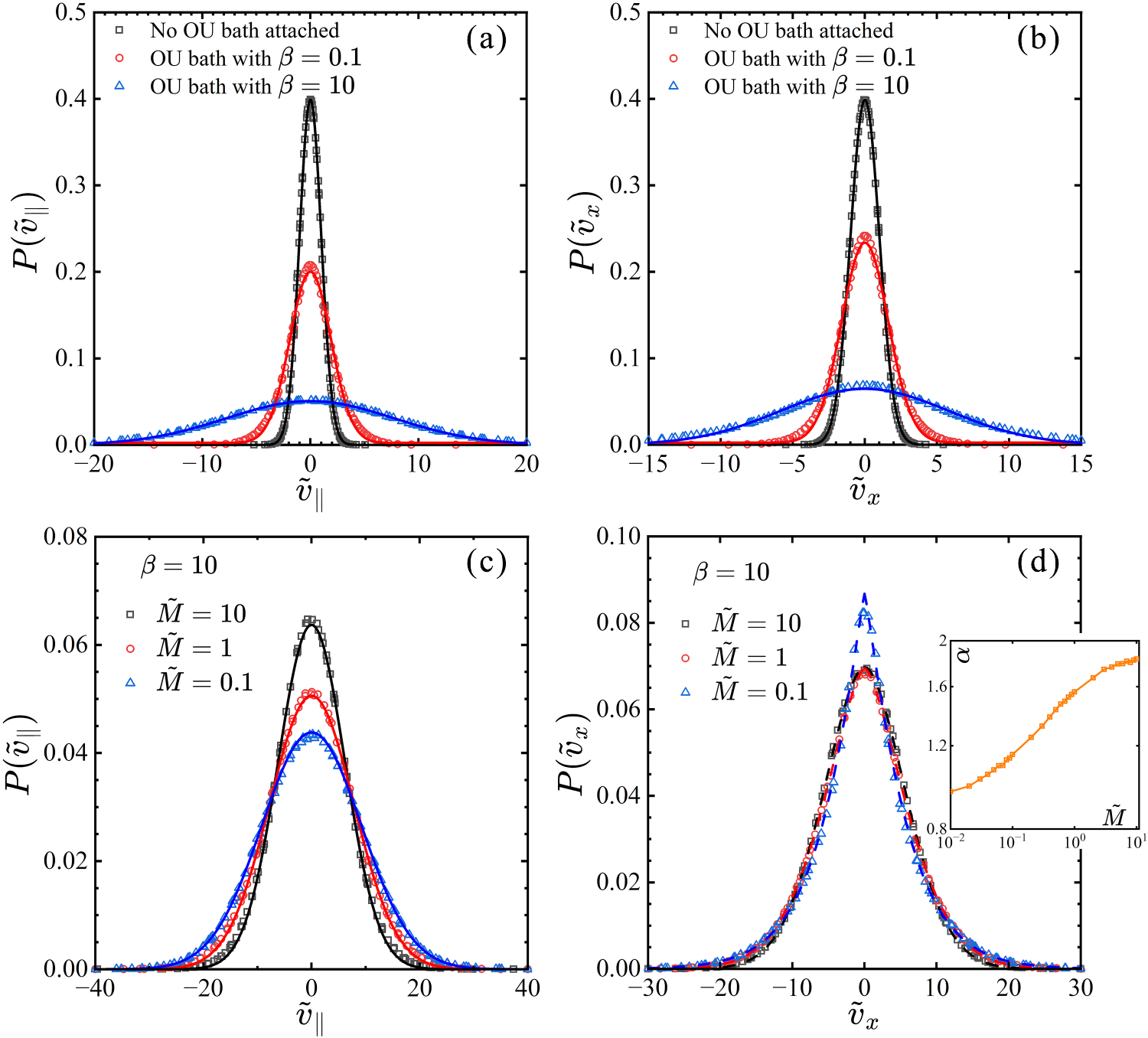}
\caption{
The velocity distribution probability of Brownian particle with/without the Ornstein-Uhlenbeck (OU) noise bath attached.
(a-b) The distribution probability of velocity of Brownian particle (BP) and active OU particle (AOUP) on the orientation direction in polar coordinate (a) and on the $x$-axis in Cartesian coordinate (b). The 'No OU bath attached' case corresponds to the Brownian particle (BP). The parameter $\beta$ is set to be 0.1 and 10, respectively. Other parameters of AOUP are $\tilde A=100$, $\tilde \tau_{\rm{a}} = 1$, and $\tilde M = 1$. The solid lines are Gaussian fits.
(c-d) The velocity distribution probability of AOUP under various inertia effects ($\tilde M$). The fixed parameters are $\beta = 10$, $\tilde A=100$, and $\tilde \tau_{\rm{a}} = 1$. The solid lines are Gaussian fits. The black, red and blue dashed lines in (d) are stretched exponential fits $P(\tilde v_x) \sim \exp({-|\tilde v_x/v_c|^{\alpha}})$ with $\alpha \sim 1.83$, $1.57$, and $1.15$, respectively. Inset: the value of $\alpha$ as a function of $\tilde M$.
}
\label{FIG:AOUP}
\end{figure*}

Although there have been various investigations dedicated for the AOUP before~\cite{nguyen2021active, caprini2021inertial}, their models introduce OU noises in each degree of freedom, which is different from our case. The AOUP in our case just integrates the OU noise on the orientation direction of particle in polar coordinate. Figure~\ref{FIG:AOUP} shows the velocity distributions of both $\tilde v_{\parallel}$ in polar coordinate and $\tilde v_x$ in Cartesian coordinate of the Brownian particle (BP, no OU bath attached case in Figs.~\ref{FIG:AOUP}(a) and (b), whose dynamics follow the Eqs.~({\ref{EQ:parallelmotion}-\ref{EQ:noises}}) with $U = \tilde U = 0$) and AOUPs with $\beta = 0.1$ and $10$, respectively.

The $P(\tilde v_{\parallel})$ and $P(\tilde v_x)$ curves of BP are both Gaussian. It is reasonable that the BP's dynamics are spatial isotropic, the $P(\tilde v_{\parallel})$ and $P(\tilde v_x)$ curves overlap with each other and follow the distribution of $P(\tilde v_{\Theta}) = (1/\sqrt{2 \pi T}) \exp{(-\tilde v^2_{\Theta} / (2T))}$ with $T$ the temperature of system, here, the subscript $\Theta$ denotes $\parallel$, $\perp$, or $\{x, y\}$ for $2$ spatial dimensions.

In general, the velocity distribution of AOUP in each degree of freedom $P(\tilde v_i)$ should be Gaussian, follows the distribution $P(\tilde v_i) = (1/\sqrt{2 \pi T^{\ast}}) \exp{(-\tilde v_i^2 / (2T^{\ast}))}$ with the effective temperature $T^{\ast} = T+ (1/k_{\rm B}) (A \gamma_{\rm t} \tau)/(\tau + \tau_{\rm{a}}/ \beta) = T [1+ \tilde A / (1+ \tilde \tau_{\rm{a}} / \beta)]$~\cite{hou2025ornstein}, if the OU noise is introduced in each independent degree of freedom in previous investigations~\cite{nguyen2021active, caprini2021inertial}. However, the distribution changes when we just introduce the OU noise in the orientation direction of the particle in our present models. The distribution of $P(\tilde v_{\parallel})$ is always Gaussian, as expected, however, the distribution $P(\tilde v_x)$ shows an inertia-effect behavior. For the particle with a strong inertial effect with a high value of $\tilde M$, its $P(\tilde v_x)$ curve roughly obeys a Gaussian distribution, while $P(\tilde v_x)$ of the particle with a weak inertial effect (i.e., decreasing the value of $\tilde M$) will deviate from the Gaussian form. Their curves can be described using a stretched exponential fitting (i.e., $P(\tilde v_x) \sim \exp({-|\tilde v_x/v_c|^{\alpha}})$), the exponential term $\alpha$ can represent the degree to which the curve deviates from a Gaussian distribution. As shown in the inset in Fig.~\ref{FIG:AOUP}(d), the smaller the $\tilde M$, the farther the $\alpha$ from 2. The feature of the AOUP in present model will predominantly determine the main feature of OUIP, which is valuable for understanding the behavior of OUIP.

\begin{figure}[htbp]
\centering
\includegraphics[width=0.3\textwidth]{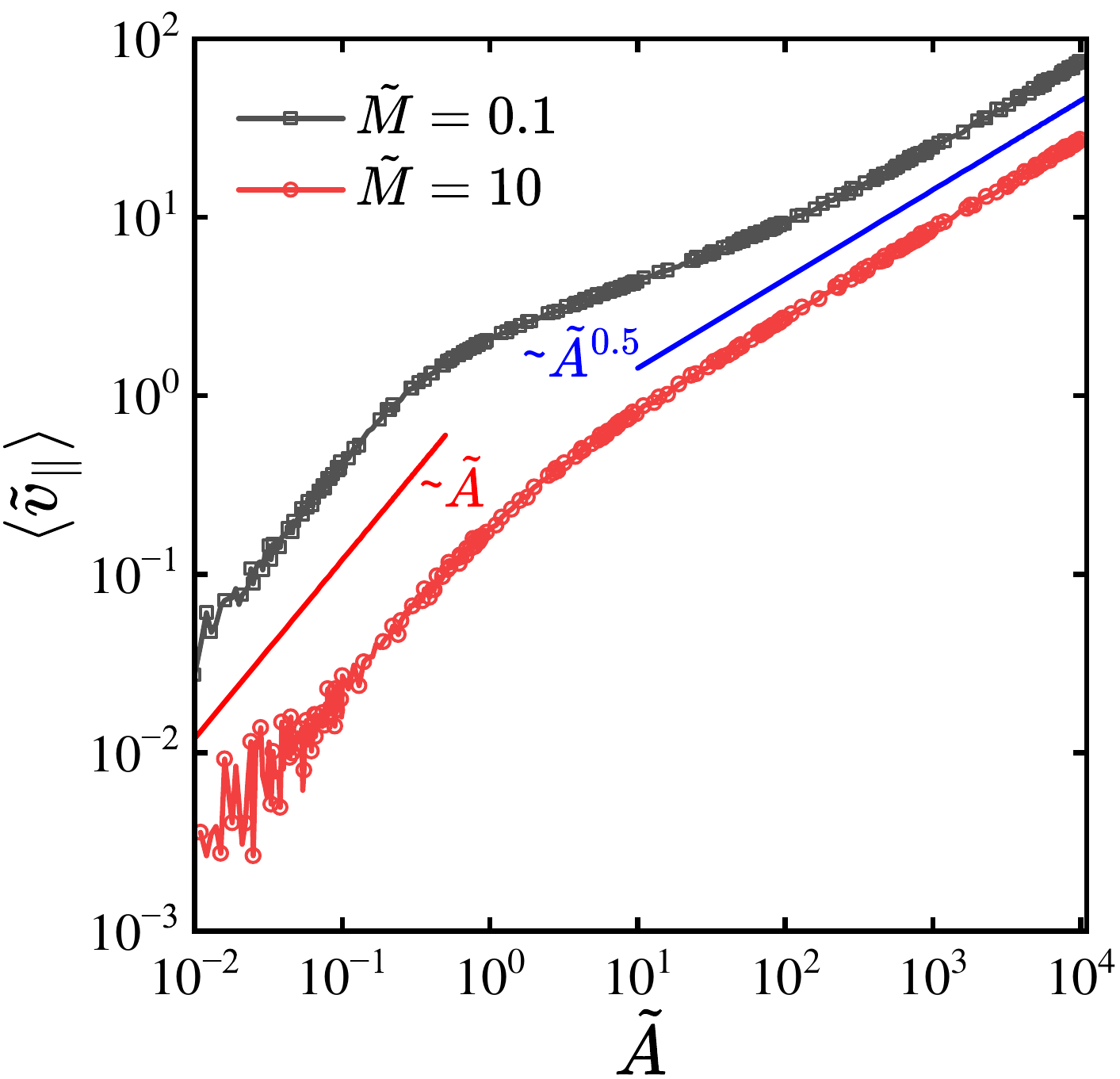}
\caption{
The obtaining net average velocity $\langle \tilde v_{\parallel} \rangle$ of OUIP as a function of $\tilde A$ under the reduced mass $\tilde M = 0.1$ and $10$, respectively.
Here, we choose $(\beta_1, \beta_2) = (10, 0.1)$, threshold velocity $\tilde v_0 = 0$, measurement time interval $\tilde \tau_{\rm{m}} = 10 \Delta \tilde t$ with $\Delta \tilde t = 0.001$, and the fixed parameter $\tilde \tau_{\rm{a}} = 1$.
}
\label{FIG:S2}
\end{figure}

\bibliography{OUIP}

\end{document}